\documentstyle[aps,epsf,psfig,multicol]{revtex}

\begin{document}
\draft
\title{Transport through cavities with tunnel barriers: a semiclassical
analysis}
\author{Markus Schreier$^1$, Klaus Richter$^2$, Gert-Ludwig Ingold$^1$,
and Rodolfo A.  Jalabert$^3$}
\address{$^1$Institut f\"ur Physik, Universit\"at Augsburg, D-86135 Augsburg,
Germany}
\address{$^2$Max-Planck-Institut f\"ur Physik komplexer Systeme, N\"othnitzer
Str.~38, D-01187 Dresden, Germany}
\address{$^3$Universit\'e Louis Pasteur, IPCMS-GEMME, 23 rue du Loess,
F-67037 Strasbourg Cedex, France}
\date{October 22, 1997}
\maketitle
\widetext
\begin{abstract}
We study the influence of a tunnel barrier on the quantum transport through
a circular cavity. Our analysis in terms of classical trajectories shows
that the semiclassical approaches developed for ballistic transport
can be adapted to deal with the case where tunneling is present.
Peaks in the Fourier  transform of the energy-dependent transmission and
reflection spectra  exhibit a
nonmonotonic behaviour as a function of the barrier height in the
quantum mechanical numerical calculations. Semiclassical analysis
provides a simple qualitative explanation of this behaviour, as well
as a quantitative agreement with the exact calculations. The experimental
relevance of the classical trajectories in mesoscopic and microwave
systems is discussed.
\end{abstract}

\raggedcolumns
\begin{multicols}{2}
\narrowtext

\section{Introduction}
\label{sec:intro}
Ballistic transport through quantum billiards has been extensively
studied in recent years due to its relevance for quantum chaos and
the possibility of physical applications. Realizations of ballistic
billiards include structured two-dimensional electron gases in semiconductor
heterostructures\cite{beena,imdat} and, exploiting the analogy between
quantum and wave mechanics, microwave cavities\cite{stoec}.
Various experiments have been designed to test theoretical ideas on
conductance fluctuations \cite{marcu,kellersurfsc,berry94}, weak localization
\cite{kelle96,chang94,lee97,taylo97} and the signatures of
classical integrability. The main theoretical tool for making the connection
between the quantum and classical properties is the semiclassical expansion
\cite{gutzw90,jal}.
This intuitive and powerful approach has been tested numerically
for the transport through circular cavities \cite{ishio95,lin95,schwi96}. In
particular, the identification of the most relevant trajectories for
transmission and reflection has been accurately demonstrated
(analogously to the relationship between the density of states and
periodic orbits of closed systems \cite{gutzw90}). Moreover, the
semiclassical approach has been extended by the inclusion of diffraction
effects at the entrance and exit of the cavities \cite{schwi96}.

In this work we further extend the applicability of semiclassical
methods in open systems to treat the case where tunneling takes place.
The modification of the trace formula in a closed system by the inclusion of
a potential step has recently been addressed for a circular billiard
\cite{bluem} in the context of ray splitting.
There the possibility of electrons entering in the region of higher potential
has to be taken into account. Our work shows that a tunnel barrier within
a cavity can very simply be incorporated in a semiclassical description
and changes the relative importance of different classical trajectories
in a non-trivial manner. Our interest in tunneling inside a cavity stems
from a fundamental point of view as well as from the fact that experiments
with a high potential barrier within a quantum dot (pacman) have
already been performed \cite{berry94a}. The inclusion of a dielectric
slab within a microwave cavity would also result in a barrier.

\begin{figure}
\begin{center}
\leavevmode
\epsfxsize=0.3\textwidth
\epsffile{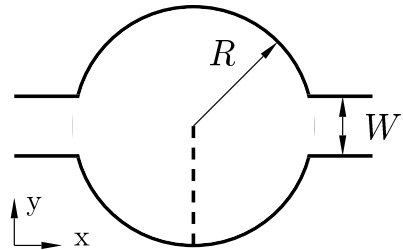}
\end{center}
\caption{The circular tunneling billiard consists of a circular billiard of
radius $R$ attached to two opposite leads of width $W$ and a thin potential
barrier of variable height shown as dashed line reaching from the edge
to the center of the circle.}
\label{fig:pacgeom}
\end{figure}

Starting from the well-studied circular billiard \cite{ishio95,lin95,schwi96},
we introduce a thin barrier placed symmetrically between the two leads and
extending from the edge to the center of the circle as shown by the dashed line
in fig.~\ref{fig:pacgeom}. The barrier height $V_b$ is variable and allows
to interpolate between a circle and the billiard studied in
ref.~\cite{berry94a}. Of special interest will be the regime where the electron
energy is of the order of the barrier height so that tunneling becomes
relevant. In the following, we therefore refer to this billiard as the
circular tunneling billiard.

We will consider phase coherent and ballistic transport through the
cavity which is attached to
two hard-wall leads of width $W$. For a fixed energy $E$ of the incident
particles there exists only a finite number of transverse modes $N$,
given by the largest integer smaller than $(E/E_0)^{1/2}$, which
contribute to the transport. Here,
\begin{equation}
E_0=\frac{\hbar^2\pi^2}{2MW^2}
\label{eq:minenergy}
\end{equation}
is the energy of the lowest transverse mode in the leads and $M$ is the mass
of the particles.

\vfill

Within the Landauer formalism\cite{beena,imdat} the two-probe
conductance $g$ through the structure is just proportional to the
total transmission coefficient $T$ at the Fermi energy $E_{\rm F}$
\begin{equation}
g=\frac{e^2}{h} \ T = \frac{e^2}{h} \ \sum_{n,m} |t_{nm}|^2 \ .
\label{eq:landauer}
\end{equation}
The transmission amplitude connecting the incoming mode
$m$ to the outgoing mode $n$ is given by the projection\cite{fishe81,stone88}
\begin{eqnarray}
G_{nm}(x',x,E_{\rm F})=&&\label{eq:projgreen}\\
&&\hskip-1.5truecm\int_0^W\int_0^W{\rm d}y'{\rm d}y\,G(x',y',x,y,E_{\rm F})
\chi_n(y')\chi_m(y)\nonumber
\end{eqnarray}
of the retarded Green function $G(x',y',x,y,E_{\rm F})$ of the structure onto
the transverse modes
\begin{equation}
\chi_m(y)=\sqrt{\frac{2}{W}}\sin\left(\frac{\pi m}{W}y\right)
\label{eq:transmode}
\end{equation}
according to
\begin{equation}
t_{nm}=\frac{{\rm i}\hbar^2}{M}(k_mk_n)^{1/2} G_{nm}(x',x,E_{\rm F}) ,
\label{eq:tmn}
\end{equation}
where we have discarded an unimportant phase factor. The longitudinal
wave vector is given by 
\begin{equation}
k_{n}=\left(\frac{2M(E-n^2E_0)}{\hbar^2}\right)^{1/2}.
\label{eq:kvec}
\end{equation}
The expression (\ref{eq:projgreen}) has to be evaluated with $x$ in the 
incoming and $x'$ in the outgoing lead.  For the amplitude of reflection 
one finds the corresponding expression
\begin{equation}
r_{nm}=-\delta_{nm}+\frac{{\rm i}\hbar^2}{M}(k_mk_n)^{1/2}
G_{nm}(x',x,E_{\rm F}).
\label{eq:rmn}
\end{equation}
with $x$ and $x'$ in the incoming lead.

The transmission and reflection amplitudes can be obtained numerically
by means of the recursive Green function method\cite{lee81,baran91}. This
method uses a discretized version of
the cavity and calculates the Green function by starting from the exact Green
function in one of the leads and successively building up the solution by means
of the Dyson equation.

Alternatively, a semiclassical approach to transport can be developed
from the semiclassical path-integral form of the Green function
leading to a transmission amplitude \cite{jal}
\begin{eqnarray}
t_{nm}&=&-\frac{\sqrt{2\pi {\rm i}\hbar}}{2W}
\sum_{s(\bar{n},\bar{m})} {\rm sgn}(\bar{n}) {\rm sgn}(\bar{m}) \sqrt
{\tilde{D}_{s}}\nonumber\\
&&\hspace{1.6truecm}\times\exp{\left(\frac{\rm i}{\hbar}
\tilde{S}_{s}(\bar{n},\bar{m},E_{\rm F})-
{\rm i} \frac{\pi}{2}{\tilde\mu}_{s}\right)} \ ,
\label{eq:scltmn}
\end{eqnarray}
given as the sum over classical trajectories $s$ between the
entrance and exit cross sections with incoming and outgoing angles
$\theta$ and $\theta^{\prime}$ such that $\sin{\theta}=\bar{m}\pi/kW$ and
$\sin{\theta^{\prime}}=\bar{n}\pi/kW$ ($\bar{m}=\pm m$, $\bar{n}=\pm n$).
The reduced action $\tilde S$ is the Legendre transform of the action integral
$S$,
\begin{equation}
\tilde S(\bar n, \bar m, E_{\rm F})=S(y_0',y_0,E_{\rm F})+
\frac{\hbar\pi}{W}(\bar m y_0-\bar n y_0') ,
\label{eq:redact}
\end{equation}
where the starting and end points of the trajectory, $y_0$ and $y_0'$
respectively, are determined by the angle quantization. For billiards
$S=kL$, where $L$ is the length of the trajectory.
The amplitude is $\tilde{D}=
(Mv|\cos{\theta^{\prime}}|)^{-1}|(\partial y/\partial \theta^{\prime})_
{\theta}|$ and $\tilde{\mu}$ is the Maslov index. A
similar expression holds for the reflection amplitude, with the
difference that now the trajectories start and end in the same lead.

Direct comparison between the semiclassical amplitudes (or the
conductance) and the exact counterparts is rather difficult since
the expansion (\ref{eq:scltmn}) includes an infinite number of
terms associated with an exponentially large number of contributing
classical trajectories. However, as in the case of the trace formula, the
validity of the semiclassical approach can be established by
identifying the Fourier components of $t_{nm}$ with classical
trajectories. This has been done in ref.~\cite{schwi96} and we
verify it in our system since it gives the starting point of our
analysis.

In Sec.~\ref{sec:identification} we present the numerical calculation
of the reflection amplitude and its interpretation in terms of
classical trajectories without and with a high barrier. In
Sec.~\ref{sec:paths} we consider the influence of a tunnel barrier
on the different trajectories contributing to the reflection and
transmission amplitudes and find a nonmonotonic behaviour.
This behaviour is modeled in
Sec.~\ref{sec:sclctb} within a modification of the
semiclassical transmission amplitudes that includes tunneling in
a very simple way. In Sec.~\ref{sec:trtot} we consider a
similar analysis for the transmission and reflection coefficients,
and show that the analysis becomes considerably more involved since
we now have to deal with pairs of trajectories. In the final
section we present our conclusions and discuss the extension of our work
to the case with magnetic field.

\section{Identification of classical trajectories}
\label{sec:identification}

\begin{figure}
\begin{center}
\leavevmode
\epsfxsize=0.45\textwidth
\epsffile{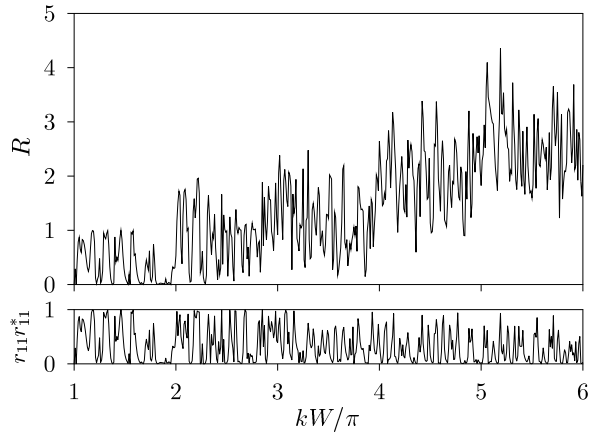}
\end{center}
\caption{Total reflection of the circular billiard and the squared modulus
of $r_{11}$ as a function of the Fermi momentum. For $1<kW/\pi<2$, both
reflection coefficients coincide as there is only one propagating mode.}
\label{fig:R}
\end{figure}

In fig.~\ref{fig:R} we present the
(exact) total reflection coefficient together with the contribution from the
lowest mode $\vert r_{11}\vert^2$. Obviously, for $1<kW/\pi<2$ the
two quantities coincide since there $N=1$.  In order to identify the classical
trajectories we carried out a discrete Fourier transformation of 600 values
of the complex reflection amplitude $r_{11}$ calculated over a momentum
interval ranging from 1 to 7 $\pi/W$. The ratio between the radius $R$ of
the circle and the width $W$ of the leads was in all calculations taken to
be $R/W=3$.

\begin{figure}
\begin{center}
\leavevmode
\epsfxsize=0.45\textwidth
\epsffile{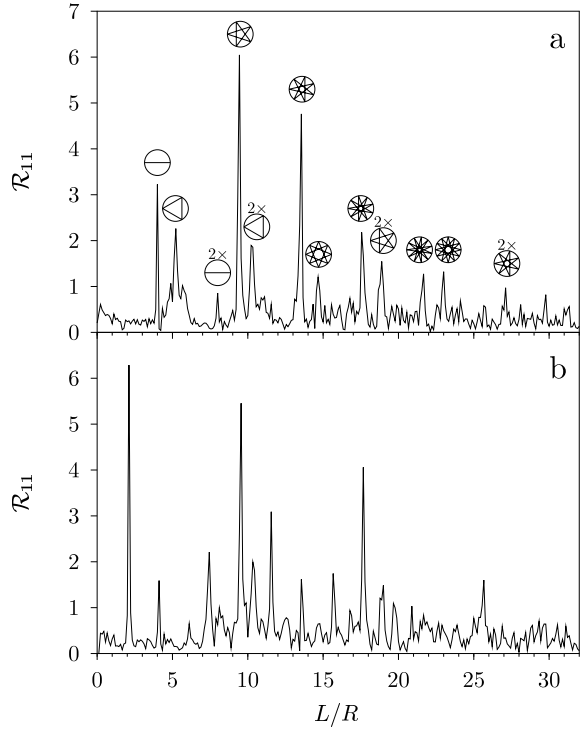}
\end{center}
\caption{Length spectrum ${\cal R}_{11}$ for (a) the case without a barrier
and (b) with an infinitely high barrier. Lengths are scaled to the radius $R$
of the circle. In the absence of a barrier we identify the peaks in the length
spectrum with periodic trajectories.}
\label{fig:r11}
\end{figure}

The semiclassical form (\ref{eq:scltmn}) of the transmission amplitude implies
that its Fourier transform with respect to the momentum $k=(2ME_{\rm F}/
\hbar^2)^{1/2}$ should exhibit peaks at lengths $L$ corresponding to the
contributing classical trajectories. Similarly, the reflection amplitudes
can be interpreted in terms of classical paths.
For a detailed analysis we now consider the power spectrum
of $r_{11}$ with respect to length. This quantity, which is shown in
fig.~\ref{fig:r11}a, is given by the squared modulus of the Fourier transform
of $r_{11}$ and will be denoted as ${\cal R}_{11}(L)$ in the following.

In complete agreement with
ref.~\cite{schwi96}, we can establish a correspondence between peaks of
${\cal R}_{11}$ and the classical trajectories including their repetitions.
The first peak is not a classical trajectory contributing to reflection, but
corresponds to diffraction off the lead mouths \cite{ishio95,schwi96}.
This effect can
be interpreted in terms of a trajectory that gets reflected back at the right
lead (ghost path). For larger lengths $L$, we can identify a triangular path,
a five-star path, a seven-star path, and so on. In agreement with the
semiclassical quantization of the initial and final angle, the star-shaped
paths are the most important ones for small mode numbers which favor the
forward direction.

The resolution of the length spectrum is limited by the width of the momentum
interval used for the Fourier transformation. While this can
easily be controlled, there are also intrinsic effects restricting the
resolution. Within a semiclassical picture, at given energy and mode numbers
the angle quantization selects the appropriate paths. Depending on the width
of the leads, the transverse position of the starting and end points in the
leads are variable and therefore a given type of trajectories exists in a
certain momentum interval.
In the Fourier transformation these trajectories will contribute with different
weights since the corresponding action will depend on the momentum, thus
yielding a finite resolution. Already the fact that a type of trajectory
effectively contributes only in a finite momentum interval may limit
the resolution more strongly than the finite interval imposed by the numerics.
In this respect it is also important to note that quantum mechanical
diffraction effects at the lead mouths \cite{ishio95,schwi96} influence the
effective momentum interval and may be relevant for the resolution.

Placing a sufficiently high barrier into the cavity yields a reflection
coefficient (not shown here) uncorrelated to that presented in
fig.~\ref{fig:R}.  On the other hand, we expect that individual trajectories
would be greatly altered by the barrier and thus we analyze the length spectrum
${\cal R}_{11}(L,V_b)$ as a function of the barrier height. For very high
barriers one can see in fig.~\ref{fig:r11}b that
new length scales have appeared rendering the identification more involved
as compared to the case of vanishing barrier.

Some features are easily explained like the appearance of a peak
at length $2R$, while in the absence of a barrier the minimum length is
$4R$. In the presence of a high barrier, the direct path may get reflected at
the barrier thus leading to a peak at half of the previous minimal length.
The peak at $4R$
now consists of two contributions, namely the direct path which
is reflected at the right lead and twice the direct path reflected at the
barrier which involves a reflection at the left lead mouth. The hierarchy
continues with a smaller peak at about $6R$ which stems from three repetitions
of the direct path reflected at the barrier.

In principle, it is not clear that an analysis in terms of classical paths
is applicable for arbitrary barrier height since tunneling necessarily
implies non-classical trajectories. However, we will show that such 
an analysis is still possible and helpful towards
the understanding of the transport problem. For instance, comparison
between figs.~\ref{fig:r11}a and b shows a large suppression of the harmonic
coming from the triangular path, while the five-star path component is
much less affected. Simple arguments given in the next section explain
this difference in behaviour.

\section{Paths in the Presence of a Tunnel Barrier}
\label{sec:paths}

For a semiclassical analysis of the energy-dependent transmission and
reflection spectra, we first have to discuss how the classical paths are
modified by the barrier. The cases where well-defined classical paths exist,
are those of the circular billiard ($V_b=0$) and the circular billiard with
a very high barrier ($V_b=\infty$). While postponing a more detailed analysis
to Sec.~\ref{sec:sclctb}, we expect that at intermediate barrier
heights, the transmission and reflection amplitudes for the billiard should be
given by both classes of trajectories properly weighted according to the
transmission and reflection coefficients of the barrier. The length spectrum
of the reflection amplitude, referred to as reflection spectrum in the
following, of the circular billiard shown in fig.~\ref{fig:r11}a displays
distinct peaks which can be associated with a triangle, a five-, and a
seven-star. In the following discussion we will focus on these three
trajectories.

\begin{figure}
\begin{center}
\leavevmode
\epsfxsize=0.45\textwidth
\epsffile{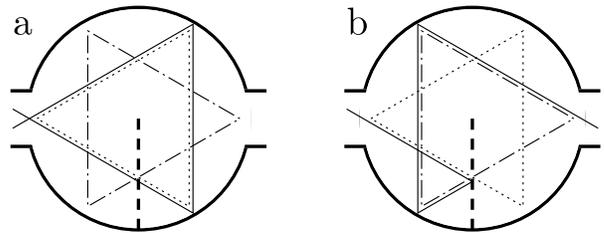}
\end{center}
\caption{The triangular path in the circular tunneling billiard may follow
the dotted triangle or its mirror image shown as dashed-dotted line.
a) Trajectory (solid line) which is transmitted at the barrier and contributes
to the reflected trajectories in the billiard. b) Trajectory (solid line)
which is reflected at the barrier and therefore contributes to the transmission
through the billiard.}
\label{fig:trianglepac}
\end{figure}

\begin{figure}
\begin{center}
\leavevmode
\epsfxsize=0.45\textwidth
\epsffile{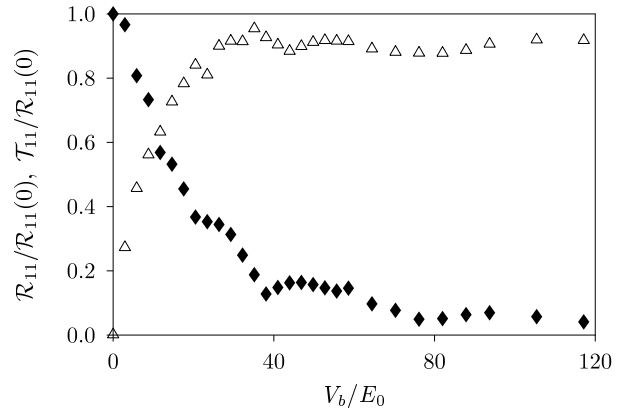}
\end{center}
\caption{Variation of the peak height in the length spectrum for the triangular
path as a function of the barrier height $V_b$. The diamonds and triangles
correspond to ${\cal R}_{11}$ and ${\cal T}_{11}$, respectively. Both
quantities are normalized with respect to ${\cal R}_{11}(0)$, i.\ e.\ the case
of vanishing barrier.}
\label{fig:3peakh}
\end{figure}

We start with the triangular path as the simplest case. In the absence of
a barrier, the trajectory just follows the triangle as shown in
fig.~\ref{fig:trianglepac}a. As the barrier height $V_b$ is increased the
transmission probability through the barrier decreases and for high barriers
the original triangle is no longer a possible path. Accordingly, the peak in
the reflection spectrum corresponding to the triangle will decrease in
amplitude with increasing barrier height. For sufficiently large $V_b$, the
possibility of reflection at the barrier has to be taken into account. As a
consequence of this reflection the path will no longer continue on the original
triangle shown as dotted line in fig.~\ref{fig:trianglepac} but follow, at
least for sufficiently thin barriers, the dashed-dotted line obtained as mirror
image with respect to a vertical line through the barrier. It is
important that this path has the same length as the original
path. However, now the end point no longer lies in the entrance lead but
in the opposite lead and thus the path reflected at the barrier will contribute
to the transmission through the billiard. Correspondingly, the triangle will
become more important in the transmission spectrum as the barrier height is
increased. This qualitative discussion is confirmed by the numerical results
for the reflection and transmission spectra, ${\cal R}_{11}$ and
${\cal T}_{11}$, shown in fig.~\ref{fig:3peakh} as diamonds and triangles,
respectively.

While the triangular path cannot appear in the reflection spectrum for very
high barriers, it is interesting to note that there is a peak at a length
corresponding to two repetitions of the triangular path. This can readily
be verified by comparing figs.~\ref{fig:r11}a and b. While one reflection
at the barrier changes the exit lead from the left to the right, an additional
reflection restores the left lead as exit lead. Therefore, the peak can be
identified with two repetitions of the path shown in
fig.~\ref{fig:trianglepac}b
including a reflection at the right lead due to diffraction at the lead
mouth.

\begin{figure}
\begin{center}
\leavevmode
\epsfxsize=0.45\textwidth
\epsffile{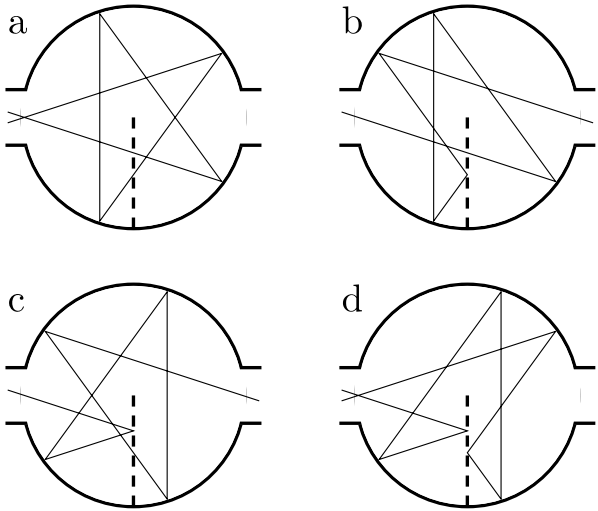}
\end{center}
\caption{Four different classes of paths related to the five-star: paths with
two transmissions or two reflections at the barrier (a and d) contribute to the
reflected trajectories in the billiard while paths with one transmission and
one reflection at the barrier (b and c) contribute to the transmission through
the billiard.}
\label{fig:fivepac}
\end{figure}

The behaviour of the five-star trajectory is more complex due to the fact that
it crosses the barrier twice. Making use of the same geometrical arguments as
for the triangle, we may distinguish four different classes
of trajectories shown in fig.~\ref{fig:fivepac} which correspond to two
transmissions at the barrier (a), one reflection and one transmission (b and
c), and two reflections (d). In fact, since the trajectory may either start
into the upper or lower\break 

\noindent
half of the billiard, these four classes correspond to eight
different trajectories which can be obtained by reading the diagrams
in the forward and backward direction.

Like for the case of the triangle, a reflection at the barrier
changes the sense of rotation in the circle thereby changing the exit lead.
As can be seen from fig.~\ref{fig:fivepac}, the paths with an even number
of reflections (a and d) contribute to the reflected paths through the billiard
while the paths with an odd number of reflections (b and c) contribute to
the transmission through the billiard.

Let us first consider the trajectories contributing to the reflection. The
trajectory (a) will only contribute for small barrier heights because it has
to be transmitted through the barrier twice. On the other hand, the trajectory
(d) will appear only for rather high barriers since it requires two reflections
at the barrier. As a consequence, we expect that the peak in the reflection
spectrum corresponding to the five-star trajectory will exhibit a minimum at
intermediate barrier heights where none of the trajectories (a) and (d)
contribute significantly. The other trajectories (b and c) appear in the
transmission spectrum at intermediate barrier heights because they have to be
reflected as well as transmitted once at the barrier.

\begin{figure}
\begin{center}
\leavevmode
\epsfxsize=0.45\textwidth
\epsffile{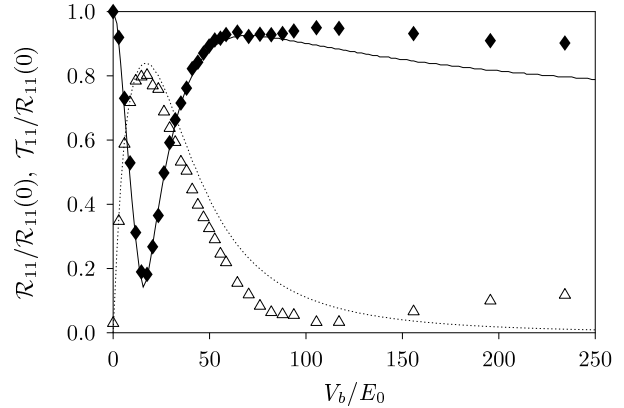}
\end{center}
\caption{Variation of the peak height in the length spectrum for the five-star
trajectory as a function of the barrier height $V_b$. The diamonds and
triangles
correspond to ${\cal R}_{11}$ and ${\cal T}_{11}$, respectively. Both
quantities are normalized with respect to ${\cal R}_{11}(0)$, i.\ e.\ the case
of vanishing barrier.  The solid and the dotted line give the results of a
semiclassical calculation taking into account tunneling through the barrier.}
\label{fig:5peakh}
\end{figure}

We conclude from the discussion of the five-star trajectory that in general
the dependence on barrier height of the peak heights in the transmission and
reflection spectra should be nonmonotonic. This is confirmed by the
numerical data shown as diamonds (${\cal R}_{11}$) and triangles
(${\cal T}_{11}$) in fig.~\ref{fig:5peakh}. The lines shown there are results
of a semiclassical analysis which will be discussed in detail in the next
section.

For this nonmonotonic behaviour it is crucial that\break 

\noindent
the reflection of
a classical path at the barrier does not change the length of
the trajectory.  Only then a change in the barrier height will not affect the
position of the peak and contributions of different paths have to be coherently
superposed.

\begin{figure}
\begin{center}
\leavevmode
\epsfxsize=0.45\textwidth
\epsffile{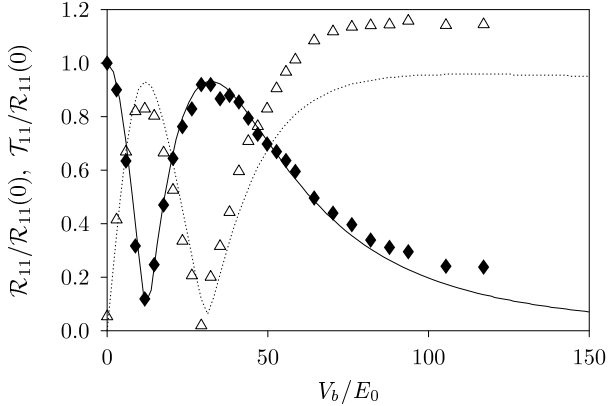}
\end{center}
\caption{Same as in fig.~\protect\ref{fig:5peakh}, but for the seven-star
trajectory.}
\label{fig:7peakh}
\end{figure}

The behaviour of the peak heights becomes more complex as the trajectories
encounter the barrier more often. This will become clear from our final
example, the seven-star trajectory. In this case, the trajectory encounters
the barrier three times giving rise to eight classes of trajectories. We may
classify these trajectories according to their behaviour at the barrier
by assigning a T or an R for each transmission or reflection, respectively.
Then, the paths contributing to the reflection spectrum are those containing
an even number of R, namely TTT, RRT, RTR, and TRR. While the first trajectory
will contribute for very small barriers, the other trajectories appear only
for sufficiently high barriers. At very high barriers none of these paths is
allowed. Accordingly, the peak in the reflection spectrum associated with the
seven-star will initially decrease with increasing barrier height, exhibit
a minimum followed by a maximum and then go to zero as the barrier height
becomes very large. This behaviour can readily be verified by comparison
with the numerical results for ${\cal R}_{11}$ shown as diamonds in
fig.~\ref{fig:7peakh}.

The behaviour of the transmission spectrum ${\cal T}_{11}$ shown in this
figure as triangles can be understood along the same line of reasoning.
The trajectories contributing to this spectrum are those with
an odd number of reflections at the barrier, i.e.\ RTT, TRT, TTR, and RRR.
Since there is at least one reflection at the barrier, seven-star trajectories
may contribute to the transmission spectrum only for finite barrier heights.
With increasing height there will be a maximum followed by a minimum 
and at\break

\begin{figure}
\begin{center}
\leavevmode
\epsfxsize=0.45\textwidth
\epsffile{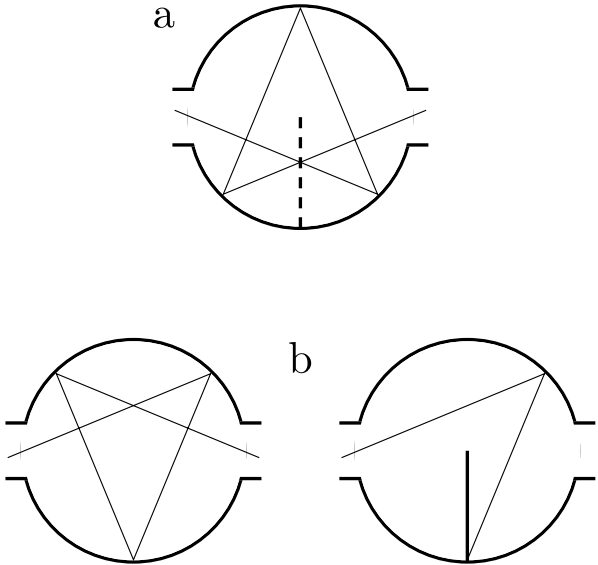}
\end{center}
\caption{Trajectory corresponding to a half of an eight-star trajectory. a)
The trajectory starting into the lower half of the billiard contributes to the
transmission through the billiard for arbitrary barrier height. b) The
trajectory starting into the upper half of the billiard contributes to the
transmission in the absence of a barrier (left) and to the reflection for
high barrier (right).}
\label{fig:eight}
\end{figure}
\noindent
very high barriers the RRR-trajectory will contribute.

So far, we have concentrated on trajectories which in the absence of a barrier
contribute to the reflection. The behaviour of classical paths connecting two
different leads, i.e.\ paths contributing to the transmission, as a function
of the barrier height is slightly more complex.  In this case, one has to
distinguish the paths first going into the upper and lower half of the
billiard. As an example, we consider the paths shown in fig.~\ref{fig:eight}
which represent one half of an eight-star. From this figure it becomes clear
that depending on the incident angle of the path there will be an even or odd
number of contacts with the barrier. Accordingly, the path shown in
fig.~\ref{fig:eight}a will behave very much like the five-star which also
encounters the barrier twice. On the other hand, the path shown in
fig.~\ref{fig:eight}b has just one contact with the barrier and its dependence
on the barrier height resembles that of the triangle. As a consequence, while
both trajectories will contribute to the transmission for low barriers, only
one of them (fig.~\ref{fig:eight}a) will do so at high barriers. The other path
will turn into a reflected path (right part of fig.~\ref{fig:eight}b)
instead.

This line of reasoning generally applies to trajectories which are transmitted
in the absence of a barrier.  Even though the analysis now becomes more
complicated, these paths also lead to nonmonotonic reflection and transmission
spectra as a function of barrier height. Again an increasing number of contacts
with the barrier will lead to an increasing number of extrema. An example will
be shown in Sec.~\ref{sec:trtot} where the spectra of the total transmission
are discussed (fig.~\ref{fig:vat}).

The previous qualitative discussion allowed us to understand the effect of
a tunnel barrier by simple consideration of classical trajectories.
In the next section we will show that a quantitative agreement with the exact
calculations can be obtained within a semiclassical approach where the
possibility of transmission or reflection at the barrier is incorporated.

\section{Semiclassical description of the circular tunneling billiard}
\label{sec:sclctb}
We now want to include tunneling into the semiclassical picture while
remaining rather close to the expression for the semiclassical Green function
(\ref{eq:scltmn}) in terms of classical paths.  To this end, we multiply the
contribution of the classical paths by amplitudes $\alpha_t$ or $\alpha_r$
accounting for each transmission or reflection of the classical path at the
barrier.

The treatment of a barrier of finite length in the circular tunneling billiard
represents a rather complicated two-dimensional problem. However, we may
approximately describe the behaviour of an electron at the barrier
as plane wave encountering a barrier of infinite length. Then the problem may
be separated into the directions perpendicular and parallel to the barrier and
the only parameter describing the scattering geometry is the incident angle
$\phi$. At this point it is important to note that the sequence of
transmissions and reflections at the barrier matters. For example the
five-star trajectories shown in figs.~\ref{fig:fivepac}b and c which, if read
from left to right, correspond
to TR and RT, respectively, have different incident angles for the transmission
and reflection events.

For an infinitely long barrier the relevant momentum component is the one
perpendicular to the barrier
\begin{equation}
k_{\bot}=k\cos(\phi).
\label{eq:k1}
\end{equation}
We now may use the standard results for one-dimensional barrier
penetration to approximate the transmission and reflection amplitudes by
\begin{eqnarray}
\alpha_{\rm t}(\phi)&=&\frac{2k_{\bot}k_{\bot}'\exp(-{\rm i}k_{\bot}b)}
{2k_{\bot}k_{\bot}'\cos(k_{\bot}' b)- {\rm i}(k_{\bot}^2+k_{\bot}'^2)
\sin(k_{\bot}' b)}\label{eq:alphat}\\
\alpha_{\rm r}(\phi)&=&{\rm i}\frac{k_{\bot}'^2-k_{\bot}^2}{2k_{\bot}
k_{\bot}'}\sin(k_{\bot}'b)\alpha_{\rm t},
\label{eq:alphar}
\end{eqnarray}
respectively, where
\begin{equation}
k_{\bot}'=\left(k_{\bot}^2-\frac{2MV_b}{\hbar^2}\right)^{1/2}
\label{eq:k2}
\end{equation}
and $b$ is the width of the barrier.  Expressions (\ref{eq:alphat}) and
(\ref{eq:alphar}) reduce to the usual one-dimensional expressions for incident
angle $\phi=0$.

We emphasize that it would not be sufficient to take the modulus of $\alpha_t$
and $\alpha_r$ since in general the contributions of different paths have to
be added up coherently. In fact, destructive interference of paths is quite
important for the interpretation of the barrier height dependence of the length
spectra. As can be seen from (\ref{eq:alphar}), the relative phase factor
between the reflected and the transmitted path is always $\pm{\rm i}$.
Since changing a reflection at the barrier into a transmission and vice versa
will change the exit lead, two classical paths going to the same exit lead
differ by their behaviour at an even number of barrier encounters.  Therefore,
the relative phase factor will be $\pm 1$. In addition, a reflection at the
barrier does not change the classical amplitude $\tilde D$, so that the two
paths either interfere perfectly constructive or destructive.

As an example we consider the minimum in the peak height of the reflection
spectrum corresponding to the five-star shown in fig.~\ref{fig:5peakh}. There
is such a pronounced minimum only because the two contributing paths, TT and RR
(cf.\ figs.~\ref{fig:fivepac}a and d), interfere destructively. On the other
hand, ${\cal R}_{11}$ is not vanishing at the minimum. This is due
to the fact that the Fourier transformation has to be taken over a finite
energy interval. Since the barrier height at which the minimum occurs is
energy-dependent, the minimum of ${\cal R}_{11}$ will be smeared out.

For finite barrier width, the phase factors appearing in the transmission and
reflection amplitudes have an additional effect. The contribution of
the barrier region to the total action of the path will depend on the
barrier height which will result in an effective change of the length of the
trajectory.  This becomes more important as the width and height of the barrier
are increased. However, for the parameters used here, the barrier is thin
enough
so that the change in length is below the resolution of the discrete Fourier
transformation. Nevertheless, the peak height is affected. This may become
important for sufficiently high barriers and cause the decrease in the
reflection spectrum of the five-star (fig.~\ref{fig:5peakh}) at large $V_b$.
For high barriers of finite width slight changes in the scattering geometry
with respect to the ideal geometry for a thin barrier may also affect the peak
height.

We now turn to a more detailed discussion of the quantum mechanical results
and those obtained from the semiclassical approach just introduced.
The data are compared in fig.~\ref{fig:5peakh} for the five-star trajectory
and in fig.~\ref{fig:7peakh} for the seven-star. In both cases the diamonds
and triangles correspond to the quantum mechanical results for ${\cal R}_{11}$
and ${\cal T}_{11}$, respectively, while the solid and dotted lines are the
corresponding semiclassical results with the modifications described above.

The expressions for the reflection and transmission amplitudes
(\ref{eq:alphat}) and (\ref{eq:alphar})
depend on the momentum component $k_{\bot}$ perpendicular to the barrier and
the height $V_b$ and width $b$ of the barrier. The geometry of the
classical path together with the Fermi energy determines $k_{\bot}$.
In the quantum calculations the barrier was implemented by increasing
on three lattice points the potential to $V_b$. For the curves shown in
figs.~\ref{fig:5peakh} and \ref{fig:7peakh} we used an effective barrier
width of 3.5 lattice spacings.

From figs.~\ref{fig:5peakh} and \ref{fig:7peakh} we find, that at not too high
barriers the agreement between the quantum mechanical data and the
semiclassical theory modified for tunneling is very good. On the other hand,
for high barriers deviations do occur. This seems to be at odds with the
fact that in the limit of infinite barrier the modified theory becomes
equivalent to the usual semiclassical expansion.

A qualitative deviation appears for rather high barriers in the transmission
spectrum of the five-star and the reflection spectrum of the seven-star
where the quantum mechanical data saturate at a finite value. This has been
checked for barriers as high as $703E_0$. On the other hand, the semiclassical
result decreases to zero since at least one barrier transmission is needed
in order to get a path which describes transmission through the billiard
and has the length of a five-star. The same holds for paths which describe
reflection at the billiard and have the same length as a seven-star.

\begin{figure}
\begin{center}
\leavevmode
\epsfxsize=0.45\textwidth
\epsffile{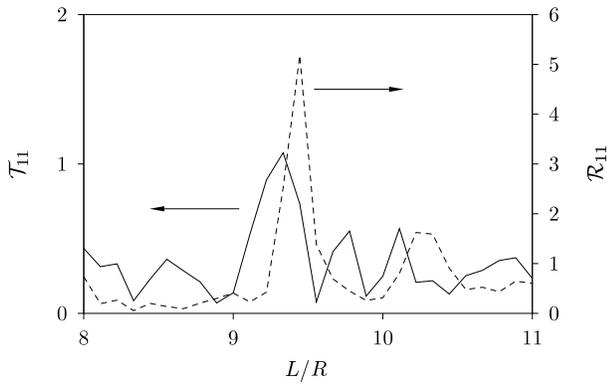}
\end{center}
\caption{Length spectra in the vicinity of the five-star peak.  The dashed
line corresponds to ${\cal R}_{11}$ in the absence of a barrier while the full
line corresponds to ${\cal T}_{11}$ for a high barrier. The scales for
${\cal T}_{11}$ and ${\cal R}_{11}$ differ by a factor of three.}
\label{fig:peakver}
\end{figure}

\begin{figure}
\begin{center}
\leavevmode
\epsfxsize=0.3\textwidth
\epsffile{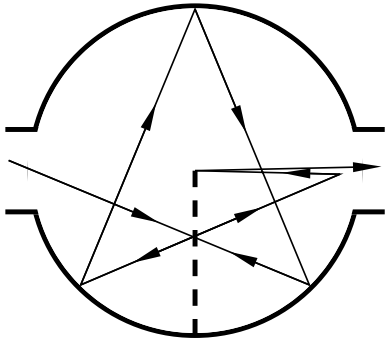}
\end{center}
\caption{Trajectory appearing in the transmission spectrum for high barriers
at a length close to that of the five-star trajectory.}
\label{fig:halfeight}
\end{figure}

This discrepancy may be
explained by paths with length close to those of the five- or seven-star.
As an example we consider the five-star for which the reflection spectrum
in the absence of a barrier and the transmission spectrum for very high
barrier are shown in fig.~\ref{fig:peakver} as dashed and solid lines,
respectively. The highest peak of the dashed line indicates the length of
the five-star. As expected, there is no peak at this position in the
transmission spectrum. However, there exists a peak at somewhat
smaller length which is broad enough to yield a contribution at the length of
the five-star. The corresponding classical trajectory is shown in
fig.~\ref{fig:halfeight}.  It involves three reflections at the barrier as well
as a reflection at the exit lead, a quantum mechanical effect \cite{schwi96}
already mentioned above.

These arguments also provide a partial explanation of the quantitative
deviations found for high barriers in the reflection spectrum of the
five-star trajectory and the transmission spectrum of the seven-star.
In addition, as discussed in the previous section the discreteness of the
Fourier transformation in combination with the change\break

\noindent
in path length as a
function of barrier height results in another source of discrepancy.

While the agreement between the quantum mechanical and
the semiclassical calculation demonstrates that for prominent peaks like the
ones corresponding to the five- and seven-star the interpretation in terms of
classical paths is possible, the deviations just discussed show that
nevertheless the identification is not necessarily straightforward and
requires a certain amount of caution. The
dependence of peaks in the length spectrum on the barrier height may be of help
in the identification since the number of extrema is related to the number
of barrier encounters of the corresponding classical trajectory.

\section{Total transmission and reflection in the circular tunneling billiard}
\label{sec:trtot}

The scattering amplitudes discussed above should be accessible to microwave
experiments where the ``Fermi energy'' can be varied by changing the input
frequency. A barrier could be introduced by placing a dielectric slab into
the cavity. On the other hand, in transport experiments on ballistic
microstructures it is the conductance which is measured. According
to the Landauer formula, eq.~(\ref{eq:landauer}), the conductance
through a cavity is proportional to the total transmission coefficient.
Recently, measurements of transport through billiards with variable Fermi
energy have been performed \cite{kelle96,zozou97} so that it appears
feasible to experimentally determine the length spectrum of the total
transmission for the circular tunneling billiard. Since the semiclassical
approaches developed so far provide a basis for the understanding of quantum
transport, we would like to analyze now the effect of a barrier on the
scattering probabilities.

The analysis for the transmission and reflection probabilities is more
difficult than for the amplitudes because taking the squared modulus makes
$\vert t_{nm}\vert^2$ depend on pairs of trajectories. Accordingly,
a Fourier transformation yields peaks at lengths which correspond to
differences between the lengths of two classical paths. Additionally,
in order to calculate the total reflection or transmission, one has to take
into account all scattering amplitudes $r_{nm}$ or $t_{nm}$ where the mode
number is restricted by the Fermi energy via the maximum mode number $N$
introduced in Sec.~\ref{sec:intro}. This will increase the number of different
paths to be considered since at higher mode numbers paths enclosing a larger
angle with the lead axis become relevant.

Moreover, when a new mode opens up it starts by being completely reflected,
and the reflection coefficient jumps by one. This staircase effect
will introduce high harmonics in the reflection coefficient which are
not related to classical trajectories. Therefore we will focus our
attention on the total transmission coefficient $T$.

\begin{figure}
\begin{center}
\leavevmode
\epsfxsize=0.45\textwidth
\epsffile{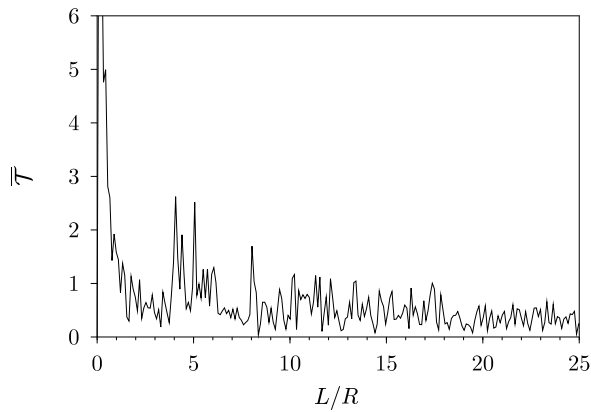}
\end{center}
\caption{Length spectrum of the total transmission $T$ for a circular billiard
without barrier.}
\label{fig:fT}
\end{figure}

As an example, the length spectrum of $T$ for the circular
billiard without barrier is shown in fig.~\ref{fig:fT}.
Due to the large amount of possible length differences the spectrum is
quite complex. Most of the peaks correspond not only to one
pair of trajectories but to a combination of several pairs of approximately
the same length difference. For example the highest peak, which is found at
$L=4.1R$, contains contributions of four different trajectories, namely a
half of an eight-star (cf.\ fig.~\ref{fig:eight}), of a twelve-star, a
sixteen-star, and a twenty-star. From these trajectories, three length
differences, $4.20R$, $4.10R$, and $4.06R$, can be constructed which are all
quite close to $L=4.1R$. In principle, higher-order stars lead to further
length differences in the same range. However, their contributions are
negligible.  Since the resolution of our discrete Fourier transformation
is $0.1R$ and the typical peak width is twice as large
(cf.\ fig.~\ref{fig:r11}), the contributions from the different combinations
of trajectories cannot be resolved.

Given the complexity of the length spectrum, we may ask the question
of whether the inclusion of a barrier can be treated as a perturbation
that simply randomizes the phases of the contributing trajectories as do
shape distortions \cite{lee97,bruusstone,marcus95} or magnetic field
changes \cite{marcu,jal}. Our analysis in the previous sections
suggests that this is not\break

\begin{figure}
\begin{center}
\leavevmode
\epsfxsize=0.45\textwidth
\epsffile{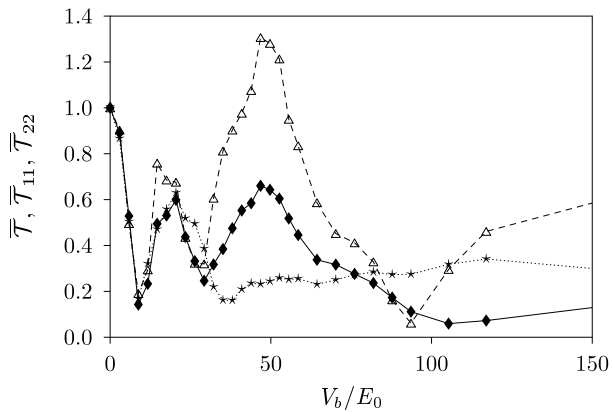}
\end{center}
\caption{Variation in the peak height of the transmission spectra
${\cal\overline{T}}$ (diamonds), ${\cal\overline{T}}_{11}$ (triangles), and
${\cal\overline{T}}_{22}$ (stars) at $L=4.1R$ as a function of the barrier
height $V_b$. The data are normalized to the respective values at $V_b=0$.
The joining lines are only guides to the eye.}
\label{fig:vat}
\end{figure}
\noindent
the case. The inclusion of a barrier has
the effect of suppressing the contribution from certain trajectories
while increasing that of the symmetry related ones, resulting in the
nonmonotonic behaviour of the peaks in the length spectrum as a
function of the barrier height which {\em cannot be interpreted as
conductance fluctuations.}

In fig.~\ref{fig:vat}  we present the dependence of the length spectrum
on the barrier height for the total transmission coefficient as well as two
individual probabilities, $|t_{11}|^2$ and $|t_{22}|^2$. In the following,
these spectra are referred to as ${\cal\overline{T}}$,
${\cal \overline{T}}_{11}$, and ${\cal \overline{T}}_{22}$, respectively.
The data are given for the peak at $L=4.1R$ to which several paths contribute
as discussed above.

The nonmonotonic behaviour resulting from the dependence of the individual
trajectories (of the contributing pairs) on $V_b$ illustrates the interplay
between semiclassics and tunneling. The fact that ${\cal\overline{T}}$ as
well as ${\cal\overline{T}}_{11}$ and ${\cal\overline{T}}_{22}$ roughly exhibit
the same structure implies that the same pairs of paths contribute to various
transmission probabilities provided the momentum interval is large enough.
Of course, predicting the peak height dependence as we did for the scattering
amplitudes is more difficult since the relative weight and the phases of the
trajectories of the contributing pairs become relevant. We therefore do not
attempt to attain the agreement of Sec.~\ref{sec:sclctb}.

Instead, we will give a qualitative explanation for the difference between
${\cal\overline{T}}_{11}$ and ${\cal\overline{T}}_{22}$.
With growing number of reflections at the circle, the incident angle of
the half-star trajectories decreases, thus suppressing the contribution
of the higher-order stars, especially to ${\cal\overline{T}}_{22}$.
Since the higher-order stars lead to more extrema in the $V_b$
dependence, the number of extrema of ${\cal \overline{T}}_{22}$ should be
smaller than that of ${\cal \overline{T}}_{11}$. This is consistent with the
numerical results of fig.~\ref{fig:vat}. The number of extrema suggests
that the dominant contribution to ${\cal \overline{T}}_{22}$ stems from
the combination of one half of an eight- and of a twelve-star. One half of
a sixteen-star should yield a contribution with a second minima for which
there is no clear indication. On the other hand, ${\cal \overline{T}}_{11}$
displays a second broad maximum which is probably a combination
of all the extrema from higher stars.

Obviously, the experimental resolution is limited and it would not
be possible to observe a length spectrum as structured as that of
fig.~\ref{fig:fT}. In mesoscopic systems temperature plays two roles. On one
hand, it controls the inelastic processes like electron-electron and
electron-phonon interactions. This results in a cut-off length
given by the inelastic mean-free path $L_{\Phi}$ beyond which no trajectory
should contribute in a semiclassical expansion. Therefore,
length differences between trajectories which are longer than $L_{\Phi}$
should not be considered, rendering in practice a length spectrum
smoother than that of fig.~\ref{fig:fT}, thus allowing for a simpler analysis.
In the language of ref.~\cite{schwi96}, the inclusion of $L_{\Phi}$
corresponds to the ``smoothing-then squaring'' process, since it is the
scattering amplitudes which loose their high harmonics.
The other effect of temperature is the rounding of the Fermi
surface that, within our semiclassical approach, cuts the large length
differences. That is, we should not consider the high harmonics of
fig.~\ref{fig:fT} beyond the thermal length $L_{T}$. Microwave cavities are
not sensitive to $L_{T}$, but have a finite $L_{\Phi}$, which can
reach very large values in state-of-the-art experiments
\cite{stoec}.

\section{Conclusions}
\label{sec:conclusions}

In this work we have studied quantum mechanically and semiclassically
the effect of a tunneling barrier on quantum transport through
ballistic cavities.  This effect is most evident in the length spectra,
i.e.\ the Fourier transform of the energy-dependent transmission and
reflection amplitudes of the cavity. We have shown that the peak heights
in the quantum mechanical length spectrum exhibit a nonmonotonic
variation upon increasing the barrier height. This behaviour is quantitatively
reproduced by combining  a semiclassical approach to conductance with
a simple model for barrier reflection and transmission of paths.
The model provides furthermore an intuitive physical picture of the
underlying process leading to the variations in the peak heights: They reflect
the superposition of coherent contributions
from paths being reflected at or transmitted through the barrier.

This mechanism is a clear-cut manifestation of tunneling orbits
in the conductance of quantum billiards, which should be in principle
observable in experiment. The effect of the tunnel barrier on individual
peaks in the {\em length} spectrum manifests itself in the energy-dependent
transmission and reflection coefficients.

We have also performed a corresponding analysis of
the effect of a tunnel barrier on the {\em area} spectrum, the
Fourier transform of the magnetic field dependent transmission at fixed
Fermi energy. However, in that case a semiclassical analysis of
the quantum mechanical results is more involved for two reasons:
Firstly, the areas of the trajectories are no longer unchanged by a
reflection at the barrier. Increasing the barrier height leads to a
shift of the spectrum to small areas \cite{lee97,berry94a}.
Secondly, in order to obtain well-resolved peaks in the area spectrum one
has to perform the Fourier transformation over a rather large range of magnetic
fields. Then, the condition that the cyclotron radius is much larger than
the system size is no longer fulfilled in our numerical calculations and the
shapes of the trajectories
become field-dependent. This leads to broadening and thus to a severe
restriction of resolution. For high magnetic fields one may even find
splitting of the peaks in the area spectrum. Hence in most cases it will be
hard to unambiguously identify peaks in the area spectrum.

Low-frequency structure has been obtained in the (non-averaged)
experimental area spectra of ballistic microstructures \cite{marcu,lee97}.
The observed peaks correspond to areas close to those of the shortest
periodic orbits. However, a clear identification between peaks and
trajectories has not been possible to establish. As discussed above the 
analysis becomes quite difficult when pairs instead of single trajectories 
are involved. Thus, a certain amount of caution has to be exerted for a 
detailed interpretation in terms of classical trajectories.

We have shown that the length spectrum admits a simple semiclassical analysis,
{\em even in the presence of tunneling}.
In view of microwave experiments
\cite{stoec} as well as recent work on cavities in two-dimensional electron
gases \cite{zozou97} which has demonstrated the possibility of measuring length
spectra, it seems feasible to experimentally verify the nonmonotonic dependence
on the barrier height discussed here.

\acknowledgements
We thank T.~Fischaleck for useful discussions.
Partial financial support was provided by DAAD and A.P.A.P.E.\
through the Procope program.

\end{multicols}
\end{document}